\documentclass[10pt,letterpaper,twocolumn]{article} 

\usepackage{ol2}
\usepackage[draft]{hyperref}
\usepackage{amsmath}

\begin{document}

\twocolumn[ 

\title{Localized mode hybridization by fine tuning of 2D random media}
\author{Laurent Labont\'{e},$^{1}$ Christian Vanneste,$^1$ and Patrick Sebbah$^{2,*}$}

\address{
$^1$ Laboratoire de Physique de la Mati\`{e}re Condens\'{e}e, Universit\'{e} de Nice-Sophia Antipolis-CNRS UMR7336,\\ Parc Valrose, 06108 Nice Cedex 02, France
\\
$^2$ Institut Langevin, ESPCI ParisTech-CNRS UMR7587, 10 rue Vauquelin, 75005 Paris, France
\\
$^*$Corresponding author: patrick.sebbah@espci.fr
}

\begin{abstract}
We study numerically interaction of spatially localized modes in strongly scattering two-dimensional media. We move eigenvalues in the complex plane by changing gradually the index of a single scatterer. When spatial and spectral overlap is sufficient, localized states couple and avoided level crossing is observed. We show that local manipulation of the disordered structure can couple several localized states to form an extended chain of hybridized modes crossing the entire sample, thus changing the nature of certain modes from localized to extended in a nominally localized disordered system. We suggest such a chain is the analog in 2D random systems of the 1D necklace states, the occasional open channels predicted by J.B. Pendry through which the light can sneak through an opaque medium.

\end{abstract}

\ocis{290.4210, 260.2710}

] 

In finite open random media, the properties of wave propagation depend on the nature of the modes. When scattering is sufficiently strong, modes may be spatially localized by disorder \cite{Anderson} if their spatial extension, given by the localization length $\xi$, is smaller than the system size $L$. Their linewidth decreases exponentially with distance from the sample boundaries \cite{Azbel,Azbel2}. In this case, wave propagation practically comes to a halt beyond a length of a few $\xi$, except at the eigenfrequencies of localized states where resonant transmission is maximal in a very narrow linewidth. In fact, extended states may occur accidentally in the very regime of localization, as predicted by J. Pendry \cite{PendryPhysC,PendryAdvPhys}. These so-called necklace states form multipeaked extended states, which span the total system. They result from the coupling of several modes which overlap both in space and in frequency. These modes are rare but they play a dominant role in transport since they are spectrally broad. They have been observed experimentally by several groups in one-dimensional (1D) random systems \cite{Wiersma,Bertolotti,Sebbah,Sebbah2,Bliokh08}. This picture was extended to larger dimensions \cite{PendryPhysC} but the existence of such chains of modes in 2D and 3D was never confirmed, probably because of their extreme scarcity. In this work, we raise the following question : can a particular random structure be designed to form necklace states instead of looking for such rare events in a collection of random realizations ? It was shown that introducing correlations in disordered structures may force the modes to extend, destroying at the same time localization \cite{Dunlap,Khule1} or conversely, improve localization
\cite{Khule2,Hui}.
 However introducing correlation is a strong perturbation which affects fundamentally the transport properties of the system.

In this Letter, we show that manipulation of the random structure at the scale of a single scatterer is sufficient to change drastically the nature of the modes of a strongly disordered system. We numerically study the spectral and spatial evolutions of localized modes in finite 2D random systems as the refractive index of a single scatterer is progressively changed. We observe mode crossing and anticrossing as modes overlap spectrally and spatially. These modes hybridize to form double-peak modes. We demonstrate that by changing the scattering index of only 2 scatterers, it is possible to couple several localized modes and tailor an extended mode, much like 1D necklace states, without changing the nature of the random system. Transmission measurements show that transport is enhanced through such an open channel.


The system is an open 2D finite structure made of 896 circular dielectric scatterers with high refractive index ($n_c=2$) embedded in a host material of index unity (vacuum) \cite{Vanneste09}. We work in the range [434.1,441.6]nm ([6.8, 6.9]$10^{14} $Hz) with scatterers radius $r=60$ nm and surface filling fraction $\Phi=40\%$. The system area is $L \times L =5\mu m \times 5\mu m$. Estimations of the localization length gives $\xi=1\mu m\ll L$. We solve Maxwell's equations using a finite element method \cite{Jin} to obtain the eigenstates and their complex eigenfrequencies, $\nu=\nu'+\text{i}\nu''$ in the TM configuration (electric field E perpendicular to the 2D propagation plane). All modes in this frequency range are found spatially localized in agreement with the estimated value of $\xi$.

\begin{figure}[h!]
\centerline{\includegraphics[width=70mm]{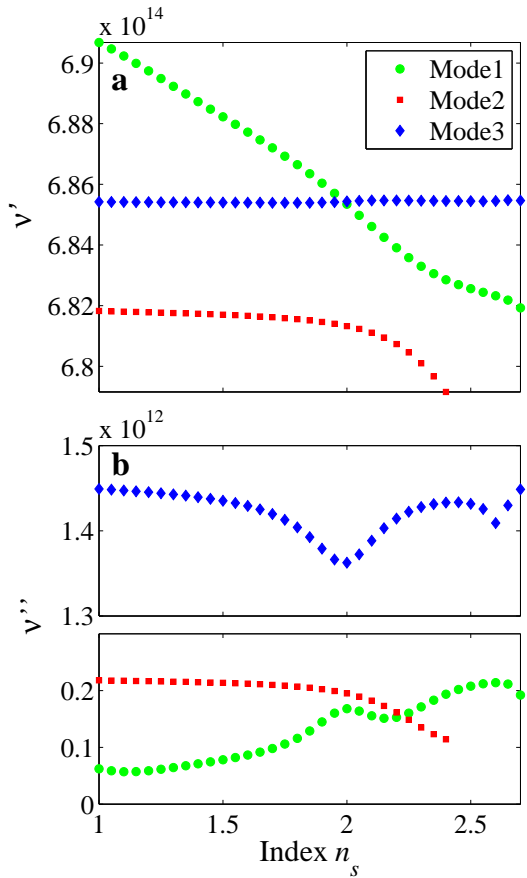}}
\caption{(Color online) Monitoring of (a) the real part $\nu'$ and (b) the imaginary part $\nu''$ of three localized states as the refractive index, $n_s$, of a particular scatterer (shown in Fig.~\ref{fig2}) is varied from 1.05 to 2.70. Y-axis has been split in (b) to span a broad interval of $\nu''$.}
\label{fig1}
\end{figure}

The refractive index $n_s$ of a single scatterer is varied from $n_s=1.05$, to $n_s=3.2$, by steps $\delta n_s=0.05$. For each value of $n_s$, all complex eigenfrequencies and eigenfunctions in the frequency range considered are computed. As they are localized away from the scatterer, most of the modes are insensitive to this local perturbation. A few modes however are shifted in frequency. The displacement of the real and imaginary parts of the eigenfrequencies can be followed "adiabatically" as the refractive index $n_s$ increases. This is seen in Fig.~\ref{fig1} where a subset of three eigenvalues, $\nu_j=\nu_j'+\text{i}\nu_j'', j \in [ 1,2,3]$, are plotted for each value of $n_s$ (named $\text{Mode}_{j \in [ 1,2,3]}$ for all values of $n_s$).

When $n_s$ increases, the real frequency $\nu'$ of $\text{Mode}_{1}$ is seen to decrease.
As $\text{Mode}_{1}$ and $\text{Mode}_{2}$ get close spectrally, level repulsion is observed for the real part of the frequency, $\nu'$. Simultaneously, the imaginary parts, $\nu''$ cross.
\begin{figure}[h!]
\centerline{\includegraphics[width=70mm]{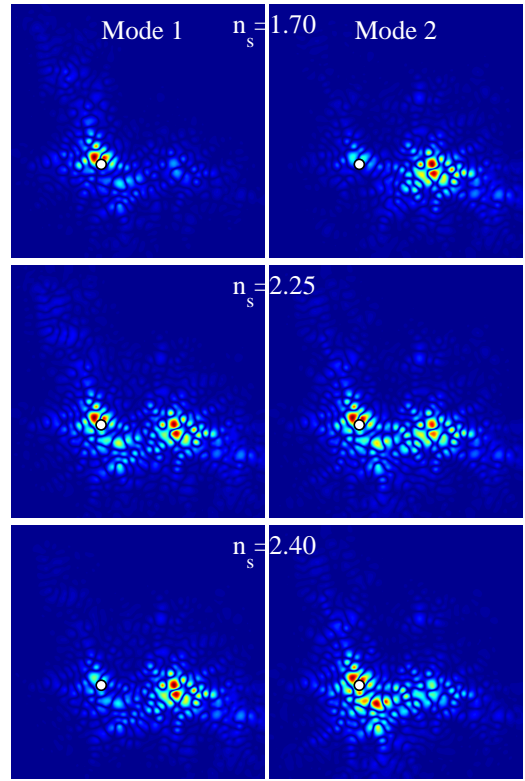}}
\caption{(Color online) Spatial distribution of the field amplitude for $\text{Mode}_{1}$ and $\text{Mode}_{2}$ of Fig.\ref{fig1} for three different values of the refractive index of the scatterer located at the white spot. Media 1 displays the evolution of the mode profiles as $n_s$ increases from 1.05 to 2.40.}
\label{fig2}
\end{figure}

The corresponding spatial distributions of their eigenfunctions are shown in Media1 and Fig.~\ref{fig2} as $n_s$ increases from 1.05 to 2.40. As $\text{Mode}_{1}$  and $\text{Mode}_{2}$  get closer to the point of avoided crossing at $n_s=2.25$, mode hybridization occurs. Two double-peaked modes are formed at 439 nm and 441 nm with a spectral width of 0.85nm. These two modes have identical amplitude maps but different phase maps (not shown).
Beyond the avoided crossing point, the spatial patterns are exchanged.
We note that the spatial distribution of $\text{Mode}_{1}$ remains unaffected as long as it is spectrally away from other modes. This indicates that the modal deformation is mainly due to mode interaction rather than to the small local perturbation.

$\text{Mode}_{1}$  and $\text{Mode}_{2}$ are modes well localized inside the system, which have small leakage through the open boundaries. Hence, like in purely Hamiltonian systems\cite{CCT}, mode interaction leads to level repulsion (anticrossing). If the system is not conservative, the energies become complex. In this case, depending on the interaction, crossing as well as anticrossing of energies can occur \cite{Brentano99}. This is true for quantum systems as well as for classical waves\cite{Brentano00}. In open random media, leakage of modes localized close to the open boundaries can become significant. Hence, crossing can also be expected. This is illustrated by the interaction between $\text{Mode}_{1}$ and $\text{Mode}_{3}$ which is a mode close to a boundary (not shown). Their real parts, $\nu'$, cross while repulsion is seen for the imaginary part, $\nu''$ (see Fig.~\ref{fig1}).

The double-peak hybridized modes seen in Fig.~\ref{fig2} for $n_s=2.25$ present an unexpected spatial profile and extension in the regime of Anderson localization, where nominal modes are exponentially localized single-peak. It was therefore tempting to manipulate further the localized modes to increase the number of hybridized localized states and try form a chain that connects one end of the system to the other. Such an extended hybridized mode would provide an open channel through which light can traverse this nominally localized random system, changing radically its transport properties. We identified modes localized on both sides of the hybridized mode of Fig.~\ref{fig2}  ($n_s$=2.25), which could open completely the channel from left to right. We found two scatterers which couple all four modes. It is remarkable that such an extended mode may coexist with localized modes. It is important to realize that the perturbation is small and that most of the modes are virtually unchanged, in contrast to e.g. correlated disorder.

\begin{figure}[h!]
\centerline{\includegraphics[width=70mm]{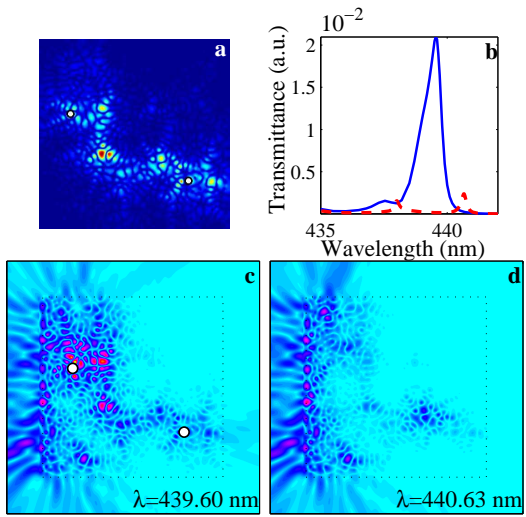}}
\caption{(Color online) (a) Hybridization of 4 localized states by setting the
index of refraction to 2.9 and 2.8 for left and right scatterers shown
as white spots, all other scatters set at $n_c = 2.0$.
(b) Transmittance spectra for plane wave impinging from left on
perturbed (full line) and unperturbed (dotted line) systems.
(c) Field amplitude spatial distribution for wave impinging on perturbed
system at $\lambda=439.60 nm$.
(d) Field amplitude spatial distribution for wave impinging on unperturbed system at $\lambda=440.63 nm$.}
\label{fig3}
\end{figure}

We probe the predominance of this beaded mode in transport by comparing in Fig.~\ref{fig3} transmission spectra through the perturbed and unperturbed systems. A plane wave is incident on the left side of the system and the total outgoing intensity at the right side is recorded as a function of the wavelength of the incident wave.

The two narrow peaks of the unperturbed system at 438 nm and 441 nm correspond to the resonant excitation of Mode 1 and Mode 2 respectively. In contrast, the perturbed system shows a higher and broader transmission near 439 nm corresponding to the excitation of the hybridized mode as shown in Fig.~\ref{fig3}. Hence, a channel has been open with enhanced broad band transmission. Further investigation is necessary to confirm the prediction that such necklace states dominate the conductance even in 2D random systems\cite{PendryViewpoint08}.


In conclusion, level repulsion was observed between two localized modes of a disordered system as the index of refraction of a single scatterer was increased. For these two modes, hybridization is observed at the anticrossing point with the formation of symmetric and antisymmetric double peaked modes. Level crossing can also occur when one mode is strongly leaking out of the sample. We show that the random structure can be manipulated further to create a chain of hybridized localized modes, extended from one end of the sample to the other.
The possibility of forming such open channels is of extreme interest. First it confirms that there is an analog in 2D random systems of the necklace states discussed in 1D random media. In 1D systems, J. Pendry predicted that these extended modes dominate the statistics of conductance. This prediction is not obvious for larger dimensions and needs more investigation. It is tempting however to make a connection between open transmission channels \cite{Imry86} and necklace structures as this was suggested by Pendry in the 1D case\cite{PendryViewpoint08}, a question to be addressed in future work.

This work was supported by the ANR under Grant No. ANR-08-BLAN-0302-01, the PACA region, the CG06, and the Groupement de Recherche 3219 MesoImage.


\begin{thebibliography}{99}

\bibitem{Anderson}
P.W. Anderson, ``Absence of Diffusion in Certain Random Lattices'', Phys. Rev. \textbf{109}, 1492 (1958).

\bibitem{Azbel}
M.Y. Azbel, ``Resonance tunneling and localization spectroscopy'', Solid State Commun. \textbf{45}, 527 (1983).

\bibitem{Azbel2}
M.Y. Azbel, ``Eigenstates and properties of random systems in one dimension at zero temperature'', Phys. Rev. B \textbf{28}, 4106 (1983).

\bibitem{PendryPhysC}
J.B. Pendry, ``Quasi-extended electron states in strongly disordered systems'',  J. Phys. C \textbf{20}, 733 (1987).

\bibitem{PendryAdvPhys}
J.B. Pendry, ``Symmetry and transport of waves in one-dimensional disordered systems'', Adv. Phys. \textbf{43}, 461 (1994).

\bibitem{Wiersma}
J. Bertolotti, S. Gottardo, and D. S. Wiersma, ``Optical Necklace States in Anderson Localized 1D Systems'', Phys. Rev. Lett. 94, 113903 (2005).

\bibitem{Bertolotti}
J. Bertolotti, M. Galli, R. Sapienza, M. Ghulinyan, S. Gottardo, L. C. Andreani, L. Pavesi, and D. S. Wiersma, ``Wave transport in random systems: Multiple resonance character of necklace modes and their statistical behavior'', Phys. Rev. E 74, 035602 (2006).

\bibitem{Sebbah}
P. Sebbah, B. Hu, J. M. Klosner, A.Z. Genack, ``Extended Quasimodes within Nominally Localized Random Waveguides'', Phys. Rev. Lett. \textbf{96}, 183902 (2006).

\bibitem{Sebbah2}
P. Sebbah, B. Hu, V. I. Kopp, and A. Z. Genack, ``Effect of absorption on quasimodes of a random waveguide'',  J. Opt. Soc. Am. B, \textbf{24}, A77 (2007).

\bibitem{Bliokh08}
K. Y. Bliokh, Y. P. Bliokh, V. Freilikher, A. Z. Genack, and P. Sebbah, ``Coupling and Level Repulsion in the Localized Regime: From Isolated to Quasiextended Modes'', Phys. Rev. Lett. \textbf{101}, 133901 (2008).

\bibitem{Dunlap}
D. H. Dunlap, H.-L. Wu, and P.W. Phillips, ``Absence of localization in a random-dimer model'', Phys. Rev. Lett. 65, 88 (1990).

\bibitem{Khule1}
U. Kuhl, F. M. Izrailev, A. A. Krokhin, and H.-J. St\"{o}ckmann, ``Experimental observation of the mobility edge in a waveguide with correlated disorder'', Appl.Phys. Lett. 77, 633 (2000).

\bibitem{Khule2}
U. Kuhl, F. M. Izrailev, and A. A. Krokhin, ``Enhancement of Localization in One-Dimensional Random Potentials with Long Range Correlations'', Phys. Rev. Lett. 100, 126402 (2008).

\bibitem{Hui}
H. Noh, J.-K. Yang, S. F. Liew, M. J. Rooks, G. S. Solomon, and H. Cao, ``Control of Lasing in Biomimetic Structures with Short-Range Order'', Phys. Rev. Lett. \textbf{106}, 183901 (2011).

\bibitem{Vanneste09}
C. Vanneste, and P. Sebbah,``Complexity of two-dimensional quasimodes at the transition from weak scattering to Anderson localization'', Phys. Rev. \textbf{A79}, 041802(R) (2009).

\bibitem{Jin}
Jin J., \emph{The finite element method in electromagnetics}(Wiley New York,1993).

\bibitem{CCT}
See for instance C. Cohen-Tannoudji, B. Diu, F. Laloe, \emph{M\'{e}canique quantique} (Herman, Paris, 1973), Ch. 4.

\bibitem{Brentano99}
P. von Brentano and M. Philipp, ``Crossing and anticrossing of energies and widths for unbound levels'', Phys. Lett. \textbf{B454}, 171 (1999) and references therein.

\bibitem{Brentano00}
M. Philipp, P. von Brentano, G. Pascovici and A. Richter, ``Frequency and width crossing of two interacting resonances in a microwave cavity'', Phys. Rev. \textbf{E62}, 1922 (2000).

\bibitem{Imry86}
Y. Imry, ``Active Transmission Channels and Universal Conductance Fluctuations'', Europhys. Lett., \textbf{1}, 249 (1986).

\bibitem{PendryViewpoint08}
J. Pendry, ``Light finds a way through the maze'', Physics \textbf{1}, 20 (2008).

\end{thebibliography}
\end{document}